\documentclass[aps,prl,twocolumn,superscriptaddress,showkeys,nofootinbib,notitlepage]{revtex4}
\usepackage[english]{babel}
\usepackage{xcolor}
\usepackage{mathtools}
\usepackage{pdfsync}
\usepackage{amssymb}
\usepackage{amsmath}

\usepackage{slashed}
\usepackage[active]{srcltx}

\def\d{\hbox{{d}\kern-.20em\hbox{l}}}

\def \matrix #1 {\left(\begin{array}{cc} #1 \end{array}\right)}

\def\II{\hbox{{1}\kern-.25em\hbox{l}}}

\newcommand{\CF}{\mathrm{T}}

\newcommand{\LL}{\mathrm{L}}

\begin{document}


\title{
 The axial-vector contributions in two-photon reactions: pion transition form factor and deeply-virtual Compton scattering at NNLO in QCD}

\author{V. M. Braun}
   \affiliation{Institut f\"ur Theoretische Physik, Universit\"at
   Regensburg, D-93040 Regensburg, Germany}

  \author{A. N. Manashov}
\affiliation{II. Institut f\"ur Theoretische Physik, Universit\"at Hamburg
   D-22761 Hamburg, Germany}
\affiliation{Institut f\"ur Theoretische Physik, Universit\"at
   Regensburg, D-93040 Regensburg, Germany}
\affiliation{
   St.Petersburg Department of Steklov
Mathematical Institute, 191023 St.Petersburg, Russia}

\author{S. Moch}
\affiliation{II. Institut f\"ur Theoretische Physik, Universit\"at Hamburg
   D-22761 Hamburg, Germany}

\author{J. Sch{\"o}nleber}
   \affiliation{Institut f\"ur Theoretische Physik, Universit\"at
   Regensburg, D-93040 Regensburg, Germany}


\begin{abstract}
Using the approach based on conformal symmetry we calculate the two-loop coefficient function for the axial-vector contributions to
two-photon processes in the $\overline{\rm MS}$ scheme. This is the last missing element for the complete next-to-next-to-leading order
(NNLO) calculation of the the pion transition form factor $\gamma^\ast\gamma\to \pi $ in perturbative QCD. The corresponding
high-statistics measurement is planned by the Belle II collaboration and will allow one to put strong constraints on the pion light-cone
distribution amplitude. The calculated NNLO corrections prove to be rather large and have to be taken into account. The same coefficient
function determines the contribution of the axial-vector generalized parton distributions to deeply-virtual Compton scattering (DVCS)
which is investigated at the JLAB 12 GeV accelerator, by COMPASS at CERN, and in the future  will be studied at the Electron Ion Collider
EIC.
\end{abstract}



\maketitle

%
%
 A wealth of data on hard exclusive reactions from a new generation of experimental facilities will become available in the
coming decade. These data are expected to have a very high precision and to provide a much deeper insight in the hadron structure as
compared to the current knowledge. A pressing question is, however, whether hard exclusive hadronic reactions are under sufficient
theoretical control to allow for fully quantitative predictions, which is highly relevant for all future high-intensity
experiments. To give an example, the $\gamma^\ast\gamma\to \pi $ transition form factor is widely regarded as the ``golden mode'' that
allows one to access the pion wave function at small transverse separations, usually referred to as the light-cone distribution amplitude
(LCDA). The measurements of this form factor at  space-like momentum transfers in the interval  $4-40\,\text{GeV}^2$ by the
BaBar~\cite{Aubert:2009mc} and Belle~\cite{Uehara:2012ag} collaborations caused  much excitement and a flurry of theoretical activity due
to the unexpected large scaling violation observed by Babar. 
At Belle II~\cite{Kou:2018nap}, the statistical uncertainty is expected to be reduced
by a factor of 8 and the total systematic uncertainty is estimated to be  at least 2 times smaller than that at Belle due to an improved
trigger efficiency. As the result, a factor 3 to 5 times more precise measurements are possible in the high-$Q^2$ region.

Another example is provided by DVCS which is an important part of the physics program at the JLAB 12 GeV upgrade~\cite{Dudek:2012vr}, is
measured also at CERN by COMPASS~\cite{Akhunzyanov:2018nut}, and in the future will be studied at the
EIC~\cite{Accardi:2012qut,AbdulKhalek:2021gbh}. This reaction is the primary source of information on the generalized parton distributions
(GPDs) of the nucleon (and, eventually, nuclei) which describe the correlation between parton's longitudinal momentum and its position in
the transverse plane. Also in this case, the accuracy of the arriving and expected data is much higher as compared to the theoretical
predictions available.

In both cases the theory framework is provided by collinear factorization and, despite obvious differences, there are some common elements.
In particular the scale dependence of the LCDAs and GPDs is governed by similar equations, and the coefficient function (CF) appearing in the
factorized expression for the $\pi \gamma^\ast\gamma $ form factor in terms of the pion LCDA is the same as the CF of the
axial GPD in DVCS. The corresponding calculations have to be advanced to NNLO accuracy that has become standard
in studies of inclusive processes and, since recently, also in semi-inclusive reactions in the framework of transverse momentum dependent factorization~\cite{Gutierrez-Reyes:2018iod}.

The new contribution of this work is a calculation of the two-loop CF for the flavor-nonsinglet axial-vector contributions in processes
with one real and one virtual photon. When combined with the three-loop anomalous dimensions calculated in \cite{Braun:2017cih}, this
result allows for the complete NNLO evaluation of the pion transition form factor and all flavor-nonsinglet contributions to DVCS. In this
letter we concentrate ourselves in the discussion of the numerical impact of the NNLO correction on $\pi\gamma^\ast\gamma$, since the DVCS
observables are more complicated and require a dedicated study.

The pion transition form factor with one real and one virtual photon, $F(Q^2)\equiv F_{\pi \gamma^*\gamma}(Q^2)$,
can be defined by the matrix element of the time-ordered product of two electromagnetic currents
\begin{align}
&\int d^4 y\, e^{iq y}\langle\pi^0(p)|T\{j^{\rm em}_\mu(y)j^{\rm em}_\nu(0)\}|0\rangle =
\notag\\
&\hspace*{3cm} =  i e^2 \varepsilon_{\mu\nu\alpha\beta} q^\alpha p^\beta F(Q^2)\,,
\label{eq:Fgamma}
\end{align}
where
$e$ is the electric charge, $p$ the pion momentum, and
$j^{\rm em}_\mu = e_u\bar u \gamma_\mu u  +  e_d\bar d \gamma_\mu d +\ldots.$
We will consider the space-like form factor, $Q^2=-q^2>0$.
The leading contribution ${\mathcal O}(1/Q^2)$ to this form factor can be written in the factorized form
\cite{Lepage:1979zb,Lepage:1980fj}
\begin{align}
  F(Q^2) = \frac{\sqrt{2}f_\pi}{6 Q^2}\!\int_{0}^{1}\! dz\, \CF(z,Q,\mu)\phi_\pi(z,\mu)\,,
\label{eq:lt}
\end{align}
where $f_\pi = 131~\text{MeV}$ denotes the pion decay constant,
$\mu$ the factorization scale,
$\CF(z,Q,\mu)$ the CF and
$\phi_\pi(z,\mu)$ the pion LCDA normalized as
$ \int_0^1dz\,\phi_\pi(z,\mu) =1$\,.

The same CF enters the axial-vector contributions to the DVCS amplitude
$\mathcal A_{\mu\nu} = g_{\mu\nu}^\perp \mathcal V + \epsilon_{\mu\nu}^\perp\mathcal A $
\begin{equation}\label{DVCS}
\hspace*{-0.2cm} \mathcal A(\xi,Q^2) = \frac12 \sum_q e_q^2 \int\limits_{-1}^1\!\frac{dx}{\xi}
 \CF\Big(\tfrac{\xi-x}{2\xi},Q,\mu\Big) \widetilde F_q(x,\xi,t,\mu)\,,
\end{equation}
where $\widetilde F_q(x,\xi,t,\mu)$ is the axial-vector GPD and $\xi$ is the skewedness parameter, see \cite{Diehl:2003ny,Belitsky:2005qn}
for details. Here it is assumed that $\CF(\tfrac{\xi-x}{2\xi},Q^2,\mu)$ is continued analytically to the $|x/\xi|>1$
region using the $\xi\to\xi-i\epsilon$ prescription.

The CF can be expanded in powers of the strong coupling, $a_s=\alpha_s(\mu)/4\pi$,
\begin{align}
\CF  &=   \CF^{(0)} + a_s \CF^{(1)} + a^2_s \CF^{(2)} +\ldots
\label{def:CF}
\end{align}
The first two terms in this series are well known~\cite{Lepage:1980fj,delAguila:1981nk,Braaten:1982yp,Kadantseva:1985kb}
\begin{align}
   \CF^{(0)}(z) &= 
   {1}/{z} + 
   {1}/{\bar z}\,,
\\
  \CF^{(1)}(z,Q,\mu) &=
  C_F \frac{1}{z} \Big[\ln^2 z -\frac{z\ln z}{\bar z } - 9
+ \left(3 +2 \ln z \right) \LL \Big]
\notag\\& \quad + (z\leftrightarrow \bar z)\,,
\end{align}
where
$ \bar z = 1-z$ and $\LL = \ln(Q^2/\mu^2)\,.$
The two-loop correction  $\CF^{(2)}$
is the subject of this work. It  can be decomposed in the
contributions of three color structures:
\begin{align}
  \CF^{(2)} &= C_F^2  \CF^{(2)}_{\rm P} + \frac{C_F}{N_c}  \CF^{(2)}_{\rm NP} + \beta_0  C_F \CF^{(2)}_{\beta}\,,
\end{align}
where $C_F= (N_c^2-1)/2 N_c$ and  $\beta_0 = 11/3 N_c-2/3 n_f$ in a $\mathrm{SU}(N_c)$ gauge theory. We obtain (for $\mu=Q$)
\begin{widetext}
\begin{align}
\CF^{(2)}_{\rm P} &= \biggl\{
\frac{2}{z} \Big( 6 \mathrm H_{0000}
- \mathrm H_{1000}
- 2 \mathrm H_{200}
- \mathrm H_{1100}
- \mathrm H_{120}
- \mathrm H_{210}
+ \mathrm H_{1110} \Big)
+ \frac{2}{\bar{z}} \mathrm H_{000}
- \frac{8}{z} \mathrm H_{100}
+ \frac{2}{\bar{z}} \mathrm H_{20}
+ \frac{4}{z} \mathrm H_{110}
\notag\\&\quad
- \left( \frac{11}{\bar{z}} + \frac{38}{3z} \right) \mathrm H_{00}
+ \frac{34}{3z} \mathrm H_{10}
+ \frac{2}{z} \zeta_2\Big( \mathrm H_{11}
- \mathrm H_2
- \mathrm H_{10}
- 4 \mathrm H_{00}\Big)
- \frac{1}{\bar{z}} \left( 2 \zeta_2 - 4 \zeta_3 - \frac{71}{2} \right) \mathrm H_0
\notag\\&\quad
+ \frac{1}{z} \left( 6 \zeta_2 + 32 \zeta_3 - \frac{64}{9} \right) \mathrm H_0
+ \frac{1}{z} \left( \frac{701}{24} + \frac{34}{3} \zeta_2 + 43 \zeta_3 + 3 \zeta_2^2 \right)  \biggr\}
+ (z \leftrightarrow \bar{z})\,,
\label{main1}
\\[2mm]
\CF^{(2)}_{\mathrm {NP}} &= \biggl\{12\left( \mathrm H_{20}
+  \mathrm H_{110}-\frac1z \mathrm H_{10}-\frac1{\bar z} \mathrm H_{0} -\zeta_3\right)
- \frac{6}{z} \mathrm H_{200}
- \left( \frac{2}{\bar{z}} + \frac{4}{z} \right)  \Big( \mathrm H_{30} + \mathrm H_{210} + \mathrm H_{31} \Big)
+ \left( \frac{2}{\bar{z}} + \frac{8}{z} \right) \mathrm H_4
\notag\\&\quad + \left( \frac{4}{\bar{z}} + \frac{6}{z} \right) \mathrm H_{22}
+ \frac{6}{\bar{z}} \Big(\mathrm H_3 - \mathrm H_{20} \Big)
+ \left[ \frac{2}{\bar{z}} \left( 1 - \zeta_2 \right) + \frac{4}{z} \left( \frac{2}{3} - \zeta_2 \right) \right] \mathrm H_{00}
- \left[ \frac{4}{z} \left( \zeta_2 - 1 \right){-\frac{1}{\bar{z}} \left( \frac{38}{3} - 2 \zeta_2 \right)} \right] \mathrm H_2
\notag\\& \quad
+ \left[ \frac{1}{z} \left( 14 \zeta_3 - \frac{32}{9} \right) + \frac{1}{\bar{z}} \left( 1- 6\zeta_2 + 6\zeta_3 \right) \right] \mathrm H_0
- \frac{1}{z} \left( 3 \zeta_2^2 + 20 \zeta_2 - 36 \zeta_3 + \frac{73}{12} \right) \biggr\}
+ (z \leftrightarrow \bar{z})\,,
\label{main2}
\\[2mm]
\CF^{(2)}_\beta &=\biggl\{-\frac{2}z \mathrm H_{000}+\frac2z\mathrm H_{100}-\frac1z \mathrm H_{110}
    +\left(\frac{10}{3z}-\frac1{\bar z}\right)\mathrm H_{00}  -\frac{14}{3z}\mathrm H_{10}
    +\left[\frac1{\bar z}\left(\frac12 +\zeta_2\right) -\frac{19}{9z}\right]\mathrm H_0
\notag\\&\quad
-\frac1z\left( \zeta_3+\frac{14}3\zeta_2+\frac{457}{24}\right)\biggr\} + (z\leftrightarrow \bar z)\,,
\label{main3}
\end{align}
\end{widetext}
where $\mathrm H_{\vec{m}} = \mathrm H_{\vec{m}}(z)$ are harmonic polylogarithms~\cite{Remiddi:1999ew}. Our result for $\CF^{(2)}_\beta$ is
in agreement with Ref.~\cite{Melic:2001wb}. The expressions for  $\CF^{(2)}_{\mathrm {P}}$ (planar), $\CF^{(2)}_{\mathrm {NP}}$ (nonplanar)
are new results. Complete results for $\CF^{(2)}(z,\mu)$ in {\tt Mathematica} format are presented in the ancillary file.

Our method of calculation is based on using conformal symmetry of QCD at critical coupling in $d=4-2\epsilon$ dimensions to relate the CF
in question to the well-known two-loop axial-vector CF in deep-inelastic scattering \cite{Zijlstra:1993sh,Moch:1999eb}.
A similar idea was used before in Ref.~\cite{Melic:2002ij} where the NNLO CF was calculated in a special, conformal renormalization scheme.
Our technique is
explained in detail in Ref.~\cite{Braun:2020yib} where the  vector CF  was calculated by the same method. The generalization to
axial-vector operators is considered in Ref.~\cite{Braun:2021tzi}. Thus in what follows we will only outline the main steps and omit
technical details.

The starting point is that QCD at the Wilson-Fischer critical point at noninteger $d=4-2\epsilon^*$ space-time dimensions, $ \epsilon^\ast
= -\beta_0 a_s - \beta_1 a_s^2 + \ldots\ $,
is a conformal theory. The strategy is to calculate the CF in conformal QCD, and go over to the ``physical'' limit $\epsilon\to 0$
at the very end, by adding a term $\sim \epsilon^\ast$:
\begin{align}
  \CF=\CF_{\ast}+\epsilon^\ast \Delta \CF\,.
\end{align}
The extra term $\Delta \CF$ requires a one-loop calculation and only affects
$\CF^{(2)}_\beta$.

\begin{figure}[t]
\includegraphics[width=0.85\linewidth]{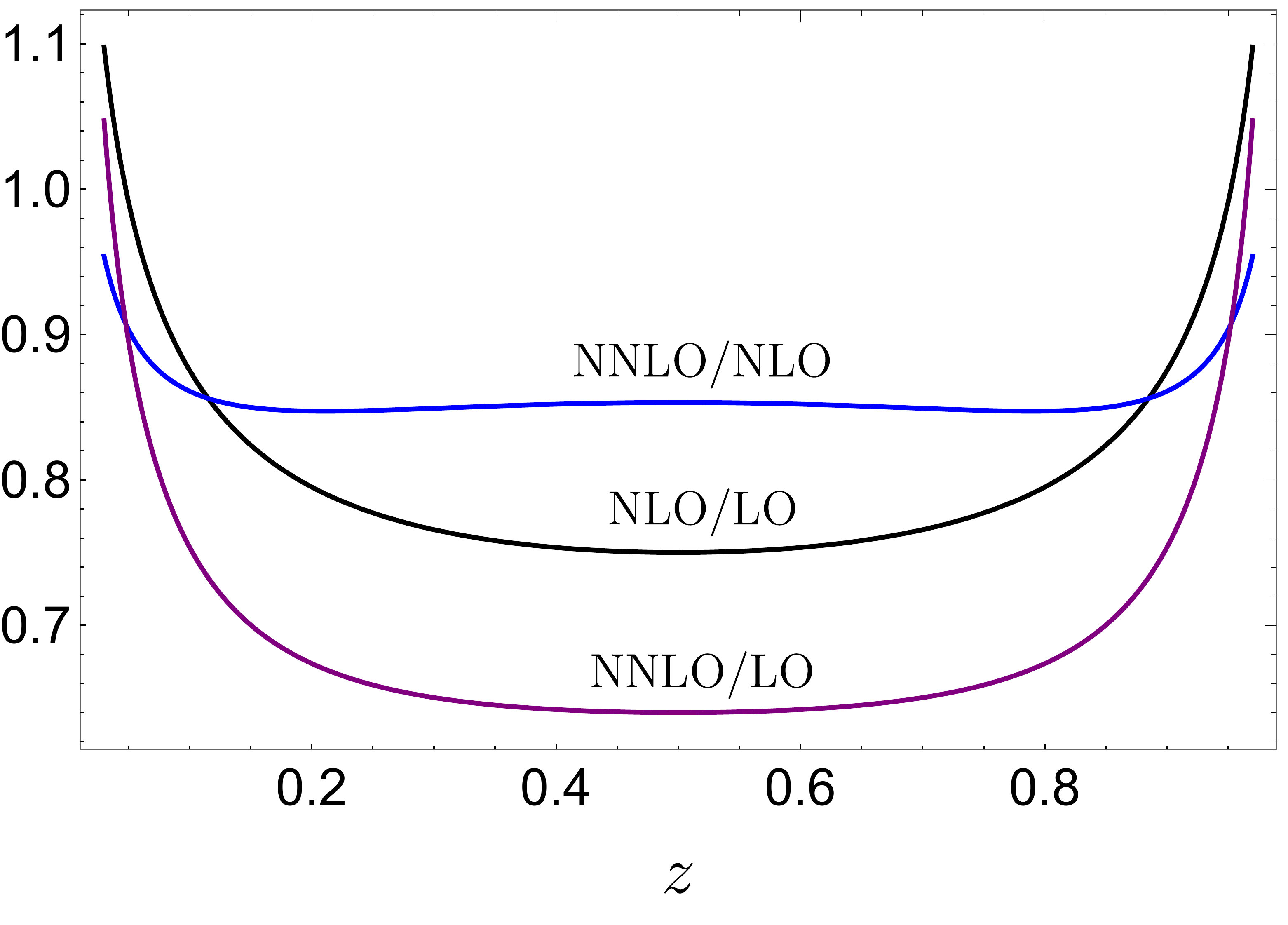}
%
\caption{The ratios $T_{\rm NLO}/T_{\rm LO}$, $T_{\rm NNLO}/T_{\rm LO}$, and $T_{\rm NNLO}/T_{\rm NLO}$
at the scale $Q^2 = \mu^2=4\, \mathrm{GeV}^2$. Here
$T_{\rm LO} = T^{(0)}$,  $T_{\rm NLO} =  T^{(0)} + a_s T^{(1)}$ and $T_{\rm NNLO} =  T^{(0)} + a_s T^{(1)}+ a_s^2 T^{(2)}$ 
are the CFs defined in \eqref{def:CF} truncated at the first (tree level), second (one-loop) and third (two-loop) terms, respectively. 
For this calculation we used $\alpha_s(\mu = 2\,\text{GeV}) = 0.3009$ and 
$n_f=4$.  
}
\label{fig:NNLO-LO}
\end{figure}

We define axial-vector operators in $d=4-2\epsilon$ dimensions using a variant of Larin's scheme~\cite{Larin:1993tq} for the $\gamma_5$ matrix,
see ref.~\cite{Braun:2021tzi}.
In $d=4$ the axial-vector operators in this scheme can be rotated to  the $\overline{\rm MS}$ scheme which is defined by the condition that
the
 vector- and  axial-vector flavor-nonsinglet operators satisfy the same evolution equation.

The CF in the $\overline{\rm MS}$ scheme can be written in the form~\cite{Braun:2020yib},
\begin{align}
 \CF_\ast &=  \CF^{(0)} \otimes K \otimes e^{\mathbb X}\otimes \mathcal{U}^{-1},
\label{scheme}
\end{align}
where  $\CF^{(0)}$ is the tree-level CF, $e^{\mathbb X}$ takes into account the conformal anomaly
\cite{Braun:2016qlg,Braun:2021tzi}, $\mathcal{U}$ is the rotation operator from  Larin's scheme to the $\overline{\rm MS}$
scheme~\cite{Braun:2021tzi}, and $K$ is a certain $\mathrm{SL}(2)$-invariant operator (i.e., $K$ commutes with the  generators of conformal
transformations). The eigenvalues of $K$ can be related to the moments of the  axial vector CF  in deep-inelastic scattering
\cite{Zijlstra:1993sh,Moch:1999eb} up to some additional factors, cf. Eq.~(3.59) in Ref.~\cite{Braun:2020yib}. We have calculated these
eigenvalues  in terms of harmonic sums  using computer algebra packages~\cite{Vermaseren:1998uu,Ablinger:2010kw,Ablinger:2014rba}. The
result satisfies the reciprocity relation \cite{Basso:2006nk,Alday:2015eya}: the asymptotic expansion of the eigenvalues of $K$ at large spin
$N$  is symmetric under the substitution $N\to -N-1$. This property provides a nontrivial test of the calculation.

Any $\mathrm{SL}(2)$ invariant operator is uniquely defined by its spectrum. Thus $K$ can be restored, and it remains to
do the convolutions in Eq.~\eqref{scheme} to obtain the final result. A direct evaluation of the
convolution in momentum space at two loops is very cumbersome, but it can be bypassed, as
explained in Ref.~\cite{Braun:2020yib}, using a position-space representation at the intermediate
step. In this way one ends up with much simpler integrals that we have calculated with the help of the {\tt HyperInt} package~\cite{Panzer:2014caa}.

The ratios of the NLO/LO, NNLO/LO and NNLO/NLO CFs at the scale $Q^2=\mu^2=4\, \mathrm{GeV}^2$
are plotted in Fig.~\ref{fig:NNLO-LO}. One sees that the two-loop $\mathcal{O}(a^2_s)$ correction has the same sign and is
roughly factor two smaller as compared to the  one-loop contribution in the bulk of the $z$ region.
The largest contribution to $\CF^{(2)}$ comes from  $\CF_\beta$ except for the endpoint regions where the leading effect
is due to the Sudakov-like double-logarithmic corrections
\begin{align*}
\CF(z,a_s)& \underset{z\mapsto 0}{\simeq}\frac1{z}\Big(1+C_F a_s \ln^2 z+\frac{(C_F a_s)^2}2\ln^4 z+
\ldots\Big).
\end{align*}
The series likely exponentiates resulting in  $ \CF(z,a_s)\sim z^{-1+a_s C_F \ln z }$.

\begin{figure}[t]
\includegraphics[width=0.85\linewidth]{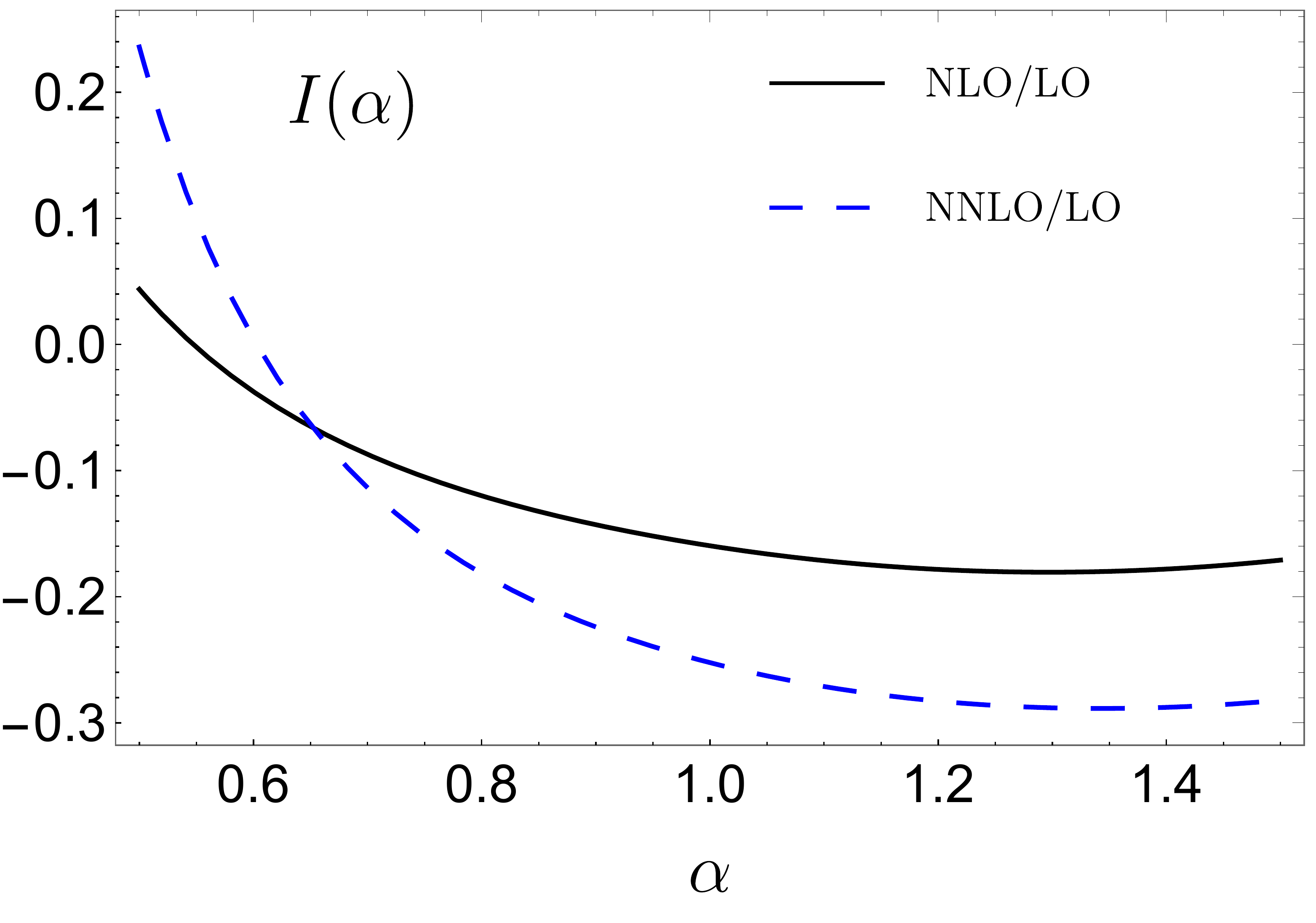}
\caption{ The NNLO/LO and NLO/LO ratios for the integral $I(\alpha)$ in Eq.~\eqref{I(alpha)} as a function of $\alpha$.}
\label{fig:AlphaPion}
\end{figure}

Note that also the sign of the correction changes, so that the resulting effect on physics
observables will depend strongly on the behavior of the parton distributions at the endpoints.
As an illustration, we show in Fig.~\ref{fig:AlphaPion} the ratios 
$I_{NLO}/I_{LO}$  and $I_{NNLO}/I_{LO}$ for the
integral
\begin{align}
 I(\alpha) = \int_{0}^{1}\! dz\, z^\alpha \bar z^\alpha \, \CF(z,Q^2,\mu^2=Q^2)
\label{I(alpha)}
\end{align}
as a function of $\alpha$ in the interval $0.5 < \alpha < 1.5$.

The pion LCDA is usually represented by the expansion in terms of
Gegenbauer polynomials which are eigenfunctions of the one-loop
evolution equations
\begin{align}
\phi_\pi(z,\mu) &= 6 z\bar z \sum_{n=0,2,\ldots} a_n(\mu) C_n^{3/2}(2z-1).
\label{Gegenbauer}
\end{align}
Here $a_0=1$ is fixed by the normalization condition, and $a_2$ is
known from lattice calculations. We use the latest result~\cite{Bali:2019dqc}
\begin{align}
  a_2(\mu_0= 2\,\text{GeV}) &= 0.116^{+19}_{-20}\,.
\label{a2}
\end{align}
For the coefficients $a_n$ with $n>2$ there exist only very weak constraints. In order to
estimate their influence, following~\cite{Bali:2019dqc},
we consider two models for the pion LCDA
(at the scale $\mu_0=2\,\text{GeV}$)
\begin{align}
 \phi^{(I)}_\pi(z,\mu_0) &= 6z(1-z)[1+ a_2(\mu_0) C^{3/2}(2z-1)]\,,
\notag\\
  \phi^{(II)}_\pi(z,\mu_0) &= B(1+\alpha,1+\alpha) z^\alpha (1-z)^\alpha\,,
\label{models}
\end{align}
where the parameter $\alpha$ in the second model is adjusted to reproduce the
same value of the second Gegenbauer coefficient $a_2$ as in Eq.~\eqref{a2}; $B(a,b)$ is Euler's $\beta$-function.

Using the expansion in Eq.~\eqref{Gegenbauer} the $\pi \gamma^\ast\gamma $ form factor to leading-twist accuracy
is given by
\begin{align}
Q^2 F(Q^2) & = {\sqrt{2} f_\pi}\sum_{n=0,2,\ldots} a_n(\mu) c_n(\mu,\mathrm{L})\,,
\end{align}
where the coefficients
\begin{align}
c_n(\mu,\LL) &= 1 + a_s c_n^{(1)}(\mu,\LL) +
 a^2_s c_n^{(2)}(\mu,\LL) + \ldots
\end{align}
are given by the Gegenbauer moments of the CF
$\CF(z,Q,\mu)$ in Eq.~\eqref{def:CF}.
The one-loop coefficients read
\begin{align}
c_{n}^{(1)} &=C_F\Big\{
4S_1^2(n\!+\!1)-\frac{4S_1(n\!+\!1)-3}{(n\!+\!1)(n\!+\!2)}+\frac{2}{(n\!+\!1)^2(n\!+\!2)^2}
\notag\\&\quad
-9
-2\left[S_1(n\!+\!2)+S_1(n)-\frac32\right]\LL\Big\},
\end{align}
where $S_1(n)$ are the standard harmonic sums, 
and the first few two-loop coefficients are given by
\begin{align}
c_0^{(2)} &= - 197.40 + 9.66 \,n_f  + \LL\, \big(68.50 -2.92\, n_f\big),
\notag\\
c_2^{(2)} &= 95.97 - 17.44\, n_f +  \LL\, \big(-178.25 + 10.9\, n_f\big)
\notag\\
&\quad
+\LL^2\, \big(45.99-1.85\, n_f \big),
\notag\\
c_4^{(2)} &= 526.97 - 43.81\, n_f +  \LL\, \big(-417.06 + 20.42\, n_f\big)
\notag\\
& \quad
+\LL^2\, \big(77.20-2.67\, n_f\big),
\notag\\
c_6^{(2)} &= 986.91 - 67.14\, n_f +  \LL\, \big(-663.97 + 27.73\, n_f\big)
\notag\\
&\quad
+ \LL^2 \,\big (101.63-3.26\, n_f\big).
\end{align}
In these expressions we included the logarithmic contributions, $\LL = \ln Q^2/\mu^2$, to allow for
the study of the factorization scale dependence. Note that the
radiative corrections (both one-loop and two-loop) to the
leading contribution $n=0$ (asymptotic LCDA) are negative,
whereas the corrections to the contributions of higher moments
are positive and increase with $n$. Thus the radiative corrections
to the form factor in general amplify the contributions of higher-order Gegenbauer polynomials
at high photon virtualities.

\begin{figure}[t]
\includegraphics[width=0.95\linewidth]{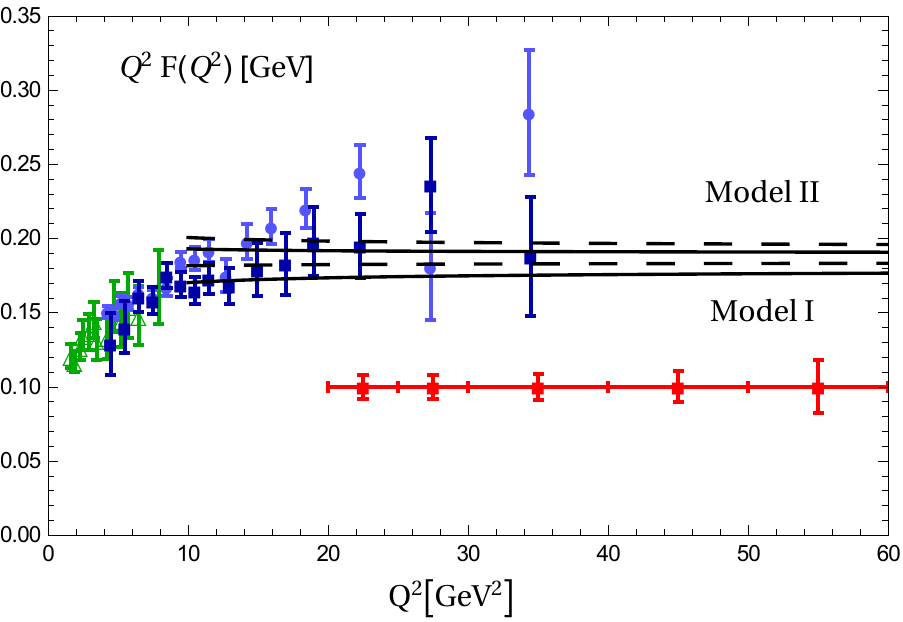}
\caption{ The  $\pi \gamma^\ast\gamma $ form factor at the NNLO (solid curves) and NLO (dashed curves)
in QCD perturbation theory for the two models of the pion LCDA in Eq.~\eqref{models}.
The experimental data are from CLEO \cite{Gronberg:1997fj} (green, open triangles),
BaBar~\cite{Aubert:2009mc} (light blue, circles) and  Belle~\cite{Uehara:2012ag} (dark blue, squares).
In addition, the expectation for the error bars achievable at Belle II~\cite{Kou:2018nap} is shown in red.
The central value for the red boxes is arbitrary.}
\label{fig:formfactor}
\end{figure}

The results for the NNLO vs. NLO calculation of the  $\pi \gamma^\ast\gamma $ form factor
for the two models of the pion LCDA in Eq.~\eqref{models} are shown in Fig.~\ref{fig:formfactor}.
In this calculation we set $\mu=Q$ and use the evolution equations both at NLO and NNLO~\cite{Braun:2017cih},
to calculate the pion LCDA \eqref{models} at this scale.
It is seen that the model dependence is
comparable in size with the projected accuracy of the Belle II measurements. The NNLO correction is about a half of the
model difference and has to be taken into account.
We conclude that the NNLO accuracy is mandatory to constrain the pion LCDA from the analysis of theory predictions with
the expected data. This, in turn, will have important
consequences on the accuracy of QCD predictions in B-decays and other hard processes with energetic pions
in the final state.

\begin{figure}[t]
\includegraphics[width=0.85\linewidth]{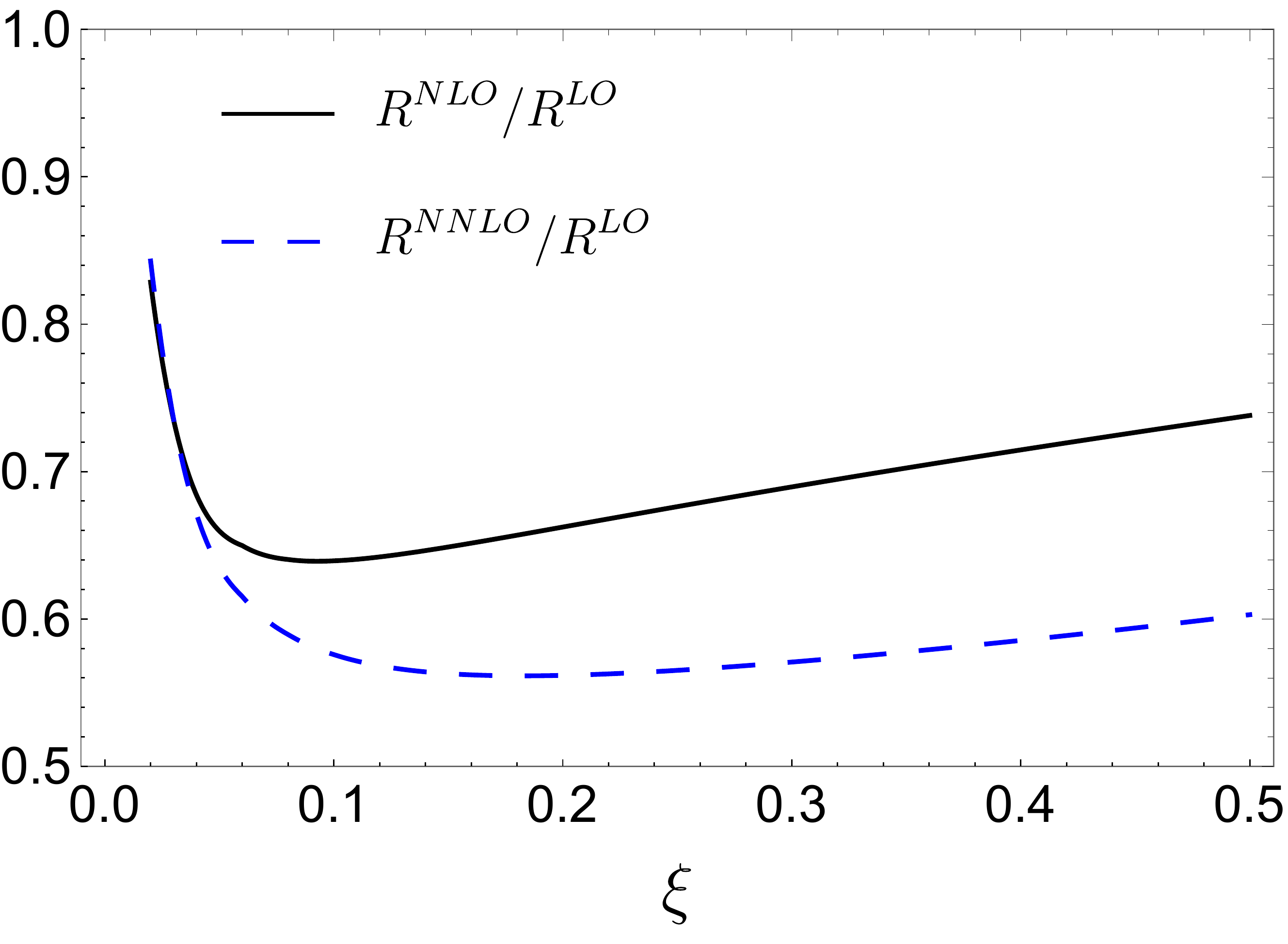}
\includegraphics[width=0.85\linewidth]{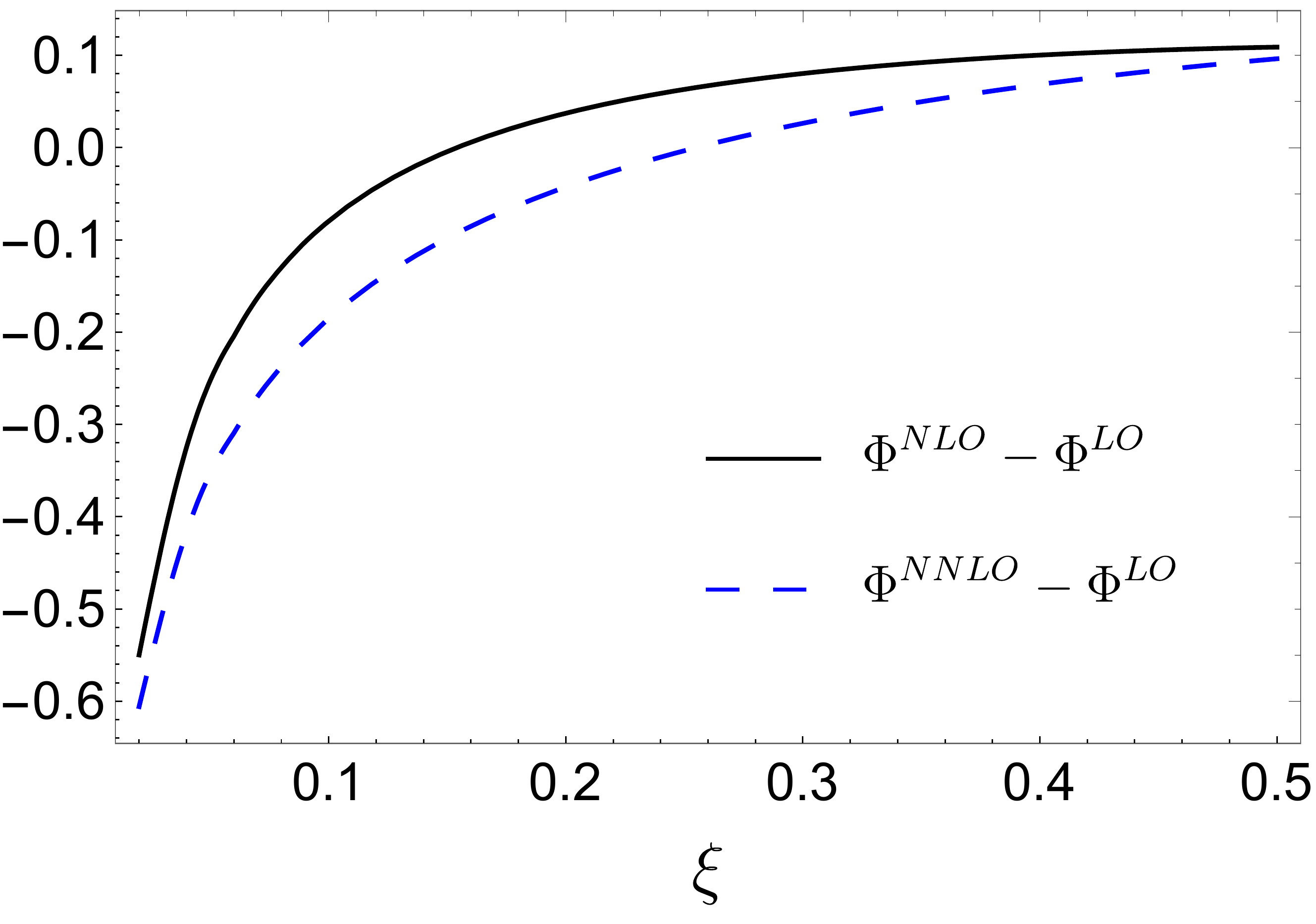}
\caption{
The axial-vector Compton form factor  $\mathcal H(\xi)$ in Eq.~\eqref{R-Phi}, calculated to NNLO and NLO accuracy.
Shown are the ratios for the absolute values (top panel) and the phase differences (bottom panel),
both with respect to the tree-level.
}
\label{fig:CFF-H}
\end{figure}

As already mentioned, the same CF enters the NNLO calculation of the contribution of the axial-vector GPD to
DVCS. The DVCS amplitude~\eqref{DVCS} is a complex-valued function.  Following \cite{Kumericki:2007sa},
we show in Fig.~\ref{fig:CFF-H} the size of the NNLO two-loop correction for the absolute value and the phase difference of the corresponding
Compton form factor at $Q^2=\mu^2=4\, \mathrm{GeV}^2$,
\begin{align}
\widetilde{\mathcal H}(\xi)= R(\xi)\, e^{i\Phi(\xi)},
\label{R-Phi}
\end{align}
for the simplest ansatz for the GPD $\widetilde H$, see Eq.~(3.330) in \cite{Belitsky:2005qn}, normalized in the forward limit
to the polarized quark density $f(\beta)\sim \beta^{1/2}(1-\beta)^4$. The results are qualitatively very similar to the
vector amplitude considered in Ref.~\cite{Braun:2020yib}: The two-loop correction is large for the absolute value of the
Compton form factor and small for the phase.
A more detailed study including the scale dependence is premature at this time, as the axial-vector GPD is practically
unknown. This task will become important in the future to analyze the forthcoming JLAB-12 and, later, EIC data.

In summary, we have used innovative computational approaches based on conformal symmetry to advance the theoretical predictions for hard
exclusive reactions to NNLO in QCD, which is the level of accuracy required by the precision of experimental data. We expect our results to
have a broad range of applications in the analyses of data from current and future high-intensity, medium energy experiments.

{\large\bf Acknowledgments:}~~~ This study was supported by DFG Research Unit FOR 2926, Grant No.~40824754, DFG grants
$\text{MO~1801/4-1}$, $\text{KN~365/13-1}$, and RSF project No.~19-11-00131. The authors are grateful to Yao Ji for communication on an
independent calculation of the same CF by another method.
We thank Sadaharu Uehara for
providing us with the estimates for the expected accuracy of the Belle II experiment.
Our special thanks are due to A.V. Pimikov for pointing out a misprint in the ancillary file.

{\large\bf Note Added:}~~~Our result for the CF in Eq.~(\ref{main1})--(\ref{main3}) 
agrees with an independent calculation of the same 
CF using a different technique~\cite{Gao:2021iqq}, cf. Eqs.~(26)--(29). 
It is easy to check   
that the analytic expressions in this work and in~\cite{Gao:2021iqq} coincide identically.


\end{document}